\documentclass{moriond}

\def\be{\begin{equation}}
\def\ee{\end{equation}}
\def\bea{\begin{eqnarray}}
\def\eea{\end{eqnarray}}

\usepackage{booktabs}
\usepackage{multirow}
\usepackage{graphicx}
\usepackage{amsmath}
\usepackage{floatrow}

\newcommand{\main}{the main paper~\cite{GWTC3TGR} }

\newcommand{\Photo}{\includegraphics[height=35mm]{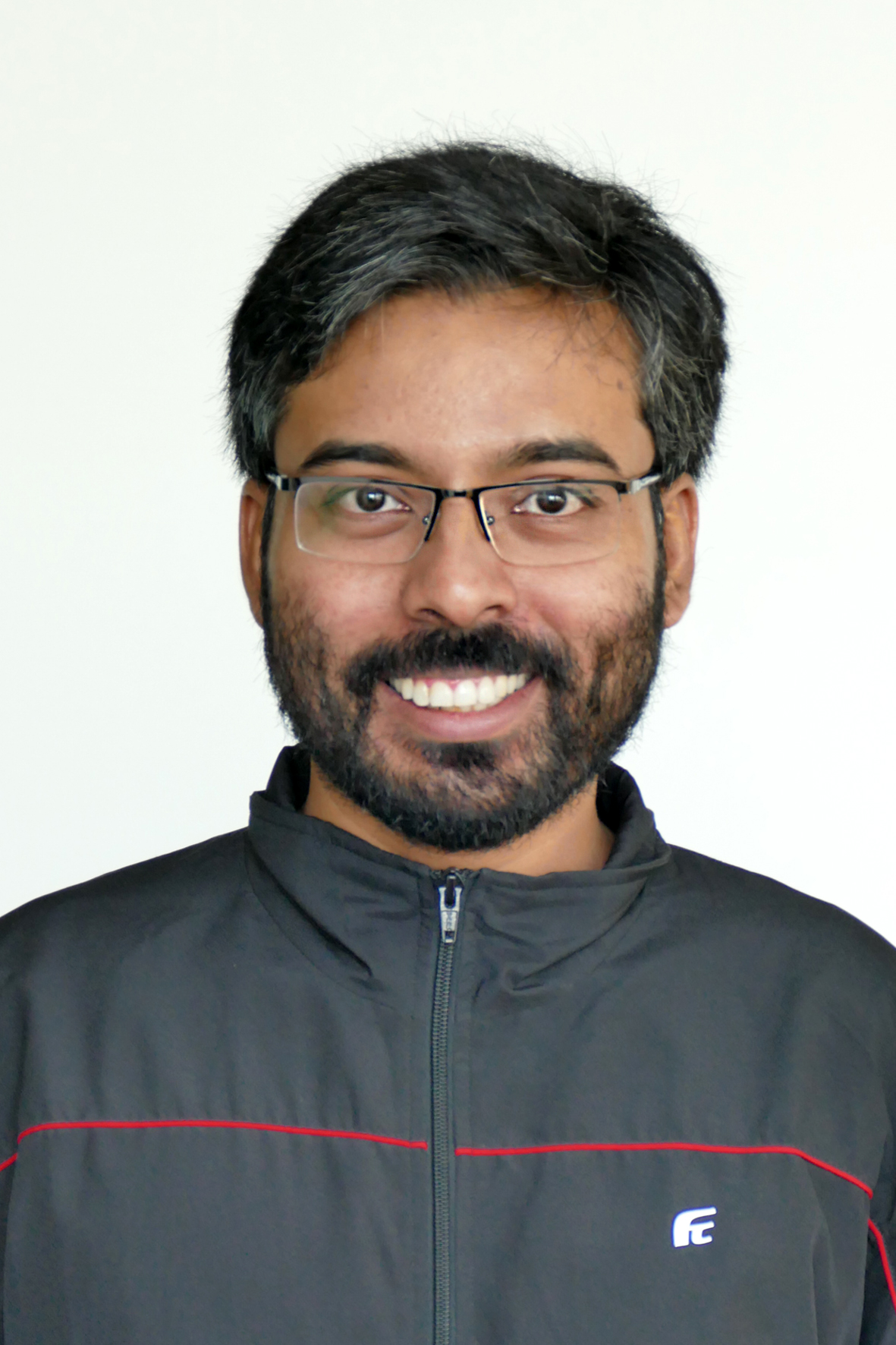}}

\begin{document}
\vspace*{4cm}
\title{Summary of Tests of General Relativity with GWTC-3}

\author{Abhirup Ghosh \\
(for the LIGO Scientific--Virgo--Kagra Collaborations)}

\address{Max Planck Institute for Gravitational Physics (Albert Einstein Institute), \\
Am M\"uhlenberg 1, Potsdam 14476, Germany}

\maketitle\abstracts{Observations of gravitational waves (GWs) by the advanced LIGO--Virgo detectors provide us with ground breaking opportunities to test predictions of Einstein's theory of general relativity (GR) in the strong field regime. In this article, we summarise the nine tests of GR performed on the new GW signals included in the third GW transient catalog, GWTC-3.  These tests include overall and self-consistency checks of the signal with the data; tests of the GW generation, propagation and polarizations; and probes of the nature of the remnant object by testing the BH ringdown hypothesis and searching for post-merger echoes. The results from the new events are combined with those previously published wherever possible. We do not find any statistically significant deviation from GR and set the most stringent bounds yet on possible departures from theory.}

\section{Introduction}
\label{sec:intro}

For more than a century, Albert Einstein's theory of general relativity (GR) has been our dominant description of gravity. The theory postulates that gravity arises as a consequence of interaction between the geometry of spacetime and its matter-energy content,  and to date, has passed all experimental challenges thrown at it~\cite{Will:2014kxa}. Observations of gravitational waves (GWs) from relativistic mergers of black holes (BHs) and neutron stars (NSs) have however, for the first time allowed us to study the nature of gravity in the highly nonlinear and dynamical regime---a regime previously inaccessible with Solar System tests,  binary pulsars or observations around supermassive BHs at centres of galaxies. Since September 2015, the Advanced LIGO-Virgo detectors~\cite{aLIGO,aVirgo} have observed 90 GW signals from mergers of such compact objects~\cite{GWTC3}, and as sensitivities have progressively improved over three separate observing runs,  GW observations have transitioned from being elusive to the routine.  

GW signals from merging binaries tracks three distinct phases of evolution. The initial \emph{inspiral}, when the two compact objects spiral in due to a backreaction of GW emission, can be well-approximated through analytic approaches like the post-Newtonian (PN)~\cite{Blanchet:2013haa} or the effective-one-body (EOB)~\cite{Buonanno:1998gg} formalisms.  Such analytic approaches, however, start to break down near the \emph{merger} where an accurate description of the signal requires solving Einstein's field equations numerically on supercomputers using techniques of numerical relativity (NR).~\cite{Sperhake:2014wpa} The final \emph{ringdown} stage, when the newly formed remnant object settles down into a stable Kerr state through GW emissions of exponentially damped sinusoids, is best described by BH perturbation theory~\cite{Kokkotas:1999bd}.  Signatures of beyond-GR physics are expected to show up as modification to binary-dynamics. However, in the absence of viable inspiral-merger-ringdown (IMR) waveform models of GWs in alternate theories of gravity, we assume our underlying signal to be well-described by GR. Any departure from this \emph{null hypothesis} would hint at the existence of a class of \emph{exotic compact objects} (ECOs), or more alarmingly, a potential breakdown of GR itself.

In the latest ``Tests of GR with GWTC-3" paper~\cite{GWTC3TGR}, henceforth referred to as ``the main paper", the LIGO Scientific--Virgo--Kagra collaborations  (LVK) use observations of GWs to test the validity of the above approaches using \emph{nine} different analyses. In the rest of this article, we list the events on which these analyses are performed in Sec.~\ref{sec:events}, details about their parameter inference in Sec.~\ref{sec:inference}, a summary of the nine tests in \main in Sec.~\ref{sec:tests}, and some concluding remarks in Sec.~\ref{sec:conclusion}. A top-level summary of these tests, their evaluation metrics and improvements of results from previous analyses is provided in Table.~\ref{tab:summary}.

\begin{table}[t]
\resizebox{\textwidth}{!}{\begin{tabular}{c c c c c c c}
\toprule
\multirow{1}{*}{Test}&\multirow{1}{*}{Section } &  \multirow{1}{*}{Quantity } & \multicolumn{1}{c}{Improvement w.r.t. GWTC-2}\\
\\
\midrule
RT &\ref{sec:con} &  $p$-value  & Not applicable
\\
IMR & \ref{sec:con} & Fractional deviation in remnant mass and spin & 1.1--1.8
 \\
PAR& \ref{sec:par} & PN deformation parameter  & 1.2--3.1
 \\
SIM & \ref{sec:sim} & Deformation in spin-induced multipole parameter  & 1.1--1.2 \\
MDR & \ref{sec:liv}& Magnitude of dispersion, $|A_{\alpha}|$ & 0.8--2.1
 \\
POL& \ref{sec:pol}& Bayes Factors between different polarization hypotheses & New Test 
\\
RD & \ref{sec:rin}&Fractional deviations in frequency (\textsc{pyRing})  & 1.1\\
 & &Fractional deviations in frequency and damping time (\textsc{pSEOB})  & 1.7--5.5
\\

ECH & \ref{sec:ech}&  Signal-to-noise Bayes Factor & New Test
\\
\bottomrule
\end{tabular}}
\caption{\label{tab:summary}
This table summarises the names of the tests performed, the corresponding sections, the metrics by which the test is quantified, and the improvement with regard to our previous analysis. The analyses performed are: RT = residuals test; IMR = inspiral--merger--ringdown consistency test; PAR = parametrised tests of GW generation; SIM = spin-induced moments; MDR = modified GW dispersion relation; POL = polarization content; RD = ringdown; ECH = echoes searches.  The last column provides the improvement in the bounds over the previous analyses reported in~\protect\cite{GWTC2TGR}. This is defined as $X_{\rm GWTC-2}/X_{\rm GWTC-3}$, where $X$ denotes the width of the 90\% credible interval for the parameters for each test, using the combined results on all eligible events.} 
\end{table}

\section{Events}
\label{sec:events}

The latest GW Transient Catalog (GWTC-3)~\cite{GWTC3} unveiled 35 new GW signals from the second half of the third observing run (O3b). Out of these 35 GW signals, \main considered only those that have significance $> 1 / 1000$ yr and were observed in at least two GW detectors. The 15 selected events, which included 14 binary black hole (BBH) events and 1 BH-NS candidate GW200115\_042309~\cite{LIGOScientific:2021qlt}, are listed in Table~\ref{tab:events} along with some of their properties.

The nine tests of GR performed in \main and outlined in Sec.~\ref{sec:tests} investigate different aspects of GW generation, propagation and polarisation. Considering the wide range of properties exhibited by the events in Table~\ref{tab:events}, not all events are suitable for all tests. The last block of columns in Table~\ref{tab:events} indicate which analyses are performed on a given event.The selection criteria include a variety of quantities including the overall signal-to-noise ratio (SNR) of a signal, the SNR in a specific portion(s) a test wants to target,  the total mass of the source or the measured spins of the compact objects.  Finally, wherever possible, we combine results of this paper with previously published results~\cite{GWTC1TGR,GWTC2TGR}.

\begin{table}
\centering
\includegraphics[width=\textwidth]{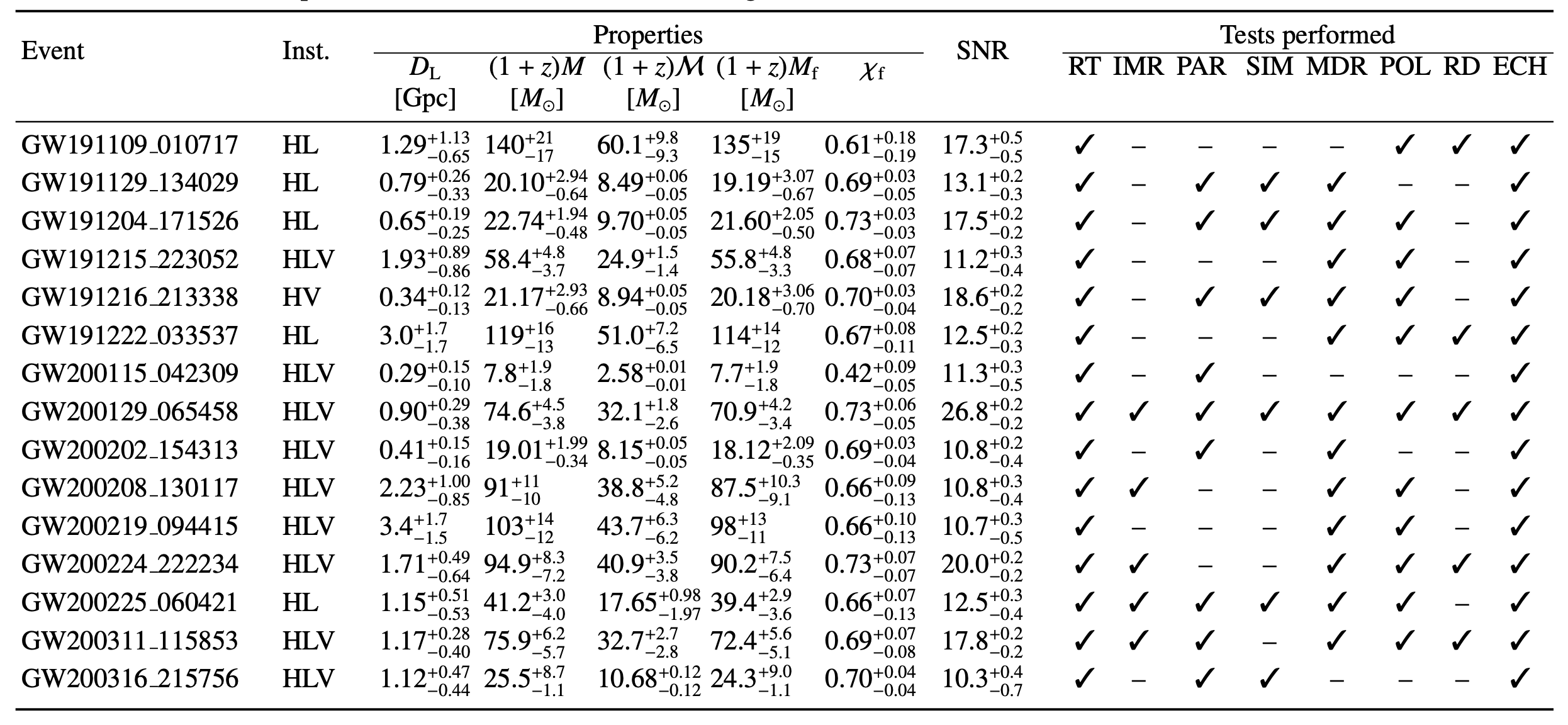}
\caption{\label{tab:events}
List of O3b events considered in \protect\main. The first block of columns gives the names of the events and the instruments (LIGO {\bf H}anford, LIGO {\bf L}ivingston, {\bf V}irgo) involved in each detection, as well as some relevant properties obtained assuming GR: luminosity distance $D_\text{L}$, redshifted total mass ($1+z)M$, redshifted chirp mass $(1+z)\mathcal{M}$, redshifted final mass $(1+z)M_\text{f}$, dimensionless final spin $\chi_\text{f}$, and network SNR. Reported quantities correspond to the median and 90\% symmetric credible intervals, as computed in Table IV in GWTC-3 \protect\cite{GWTC3}.  The last block of columns indicates which analyses are performed on a given event according to the selection criteria outlined in Sec.~\ref{sec:events}  and ~\ref{sec:tests}. }
\end{table}

\section{Parameter inference}
\label{sec:inference}

The analyses in \main assume that the underlying GW signal is well described by two families of waveform models as a BBH signal -- (aligned) spinning effective-one-body (EOB) models called \emph{SEOB}~\cite{Bohe:2016gbl,Purrer:2015tud,Cotesta:2018fcv} and phenomenological precessing waveforms called \emph{Phenom}~\cite{Hannam:2013oca,Pratten:2020ceb,Pratten:2020fqn,Garcia-Quiros:2020qpx}. Both families include models with and without higher order moments of gravitational radiation. For the BH-NS candidate GW200115\_042309~\cite{LIGOScientific:2021qlt}, matter effects of the NS are assumed to be negligible (because of the asymmetry of the binary masses) and the signal is also assumed to be well-described as the BBH signal. Using these waveform models we proceed to infer the parameters of the source using a range of statistical tools encoded in the LIGO Algorithms Library~\cite{lalsuite} ---~\textsc{LALInference}~\cite{Veitch:2014wba,lalsuite},  \textsc{Bilby}~\cite{Ashton:2018jfp,Romero-Shaw:2020owr}, \textsc{pyRing}~\cite{pyRing} and  \textsc{\textsc{BayesWave}}~\cite{Cornish:2014kda,Littenberg:2014oda}. Finally, wherever possible, \main combines information from multiple events to produce the strongest bounds possible. In this regard, we use two approaches: a restrictive approach, which assumes that these deviations appear identically across all events independent of their source properties~\cite{Zimmerman:2019wzo}, and a conservative hierarchical approach which assumes that the deviation parameters are not identical but rather belong to some underlying Gaussian population~\cite{Isi:2019asy}.

\section{Summary of Tests}
\label{sec:tests}

The tests of GR performed in  \main belong to two main theory-agnostic classes: 1) \emph{consistency tests} which search for possible violations of GR by comparing the signal or portions of the signal to the data without invoking any parametrization of deviations, and 2) \emph{parametrised tests} which introduce deviations from GR at the level of a gravitational waveform and use data to bound or constrain these beyond-GR parameters. 

\subsection{Consistency tests}
\label{sec:con}

There are two consistency tests considered in \main. The first one, called the \emph{residuals test}~\cite{LIGOScientific:2016aoc,GWTC1TGR,GWTC2TGR}, checks for the overall consistency of the signal with the data. We measure the coherent residual SNR in the data after subtracting the best-fit GR waveform for all 15 events using the \textsc{BayesWave} software, and test whether it is consistent with detector noise. We measure the distribution of SNRs in segments of noise immediately adjacent to the event, and compute the p-values for the residuals to be consistent with this background distribution. The presence of coherent noise in the residuals would indicate an inconsistency between the signal present in the data and the GR template used.  However, we find that the residuals are consistent with our understanding of detector noise.

\begin{figure}
  \begin{minipage}[l]{0.5\textwidth}
    \includegraphics[width=\textwidth]{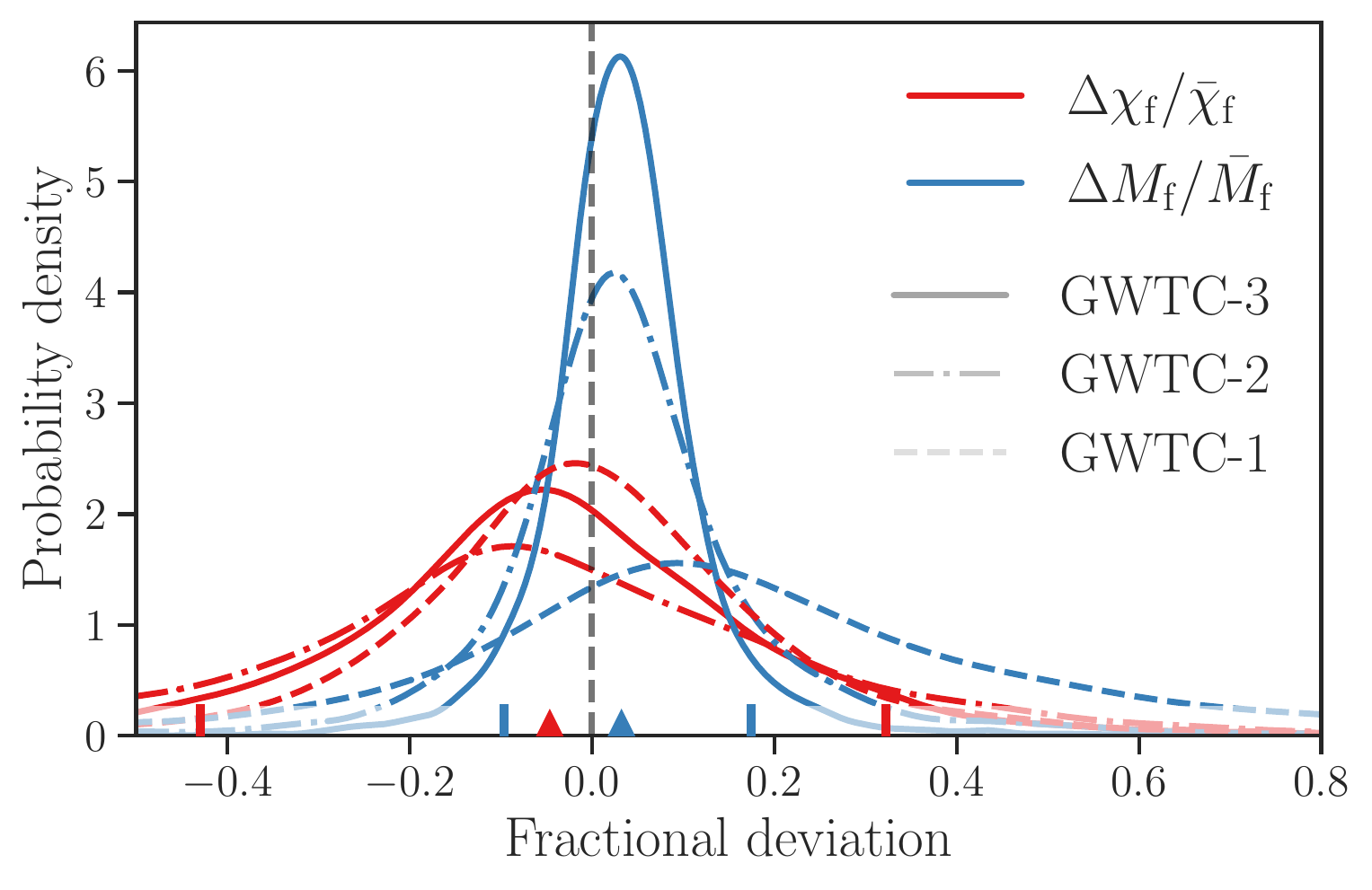}
  \end{minipage}\hfill
  \begin{minipage}[l]{0.5\textwidth}
    \caption{Distributions on the remnant mass (blue) and spin (red) fractional deviation parameters obtained by hierarchically combining the GWTC-3 events (solid trace).  For comparison, we also show the results obtained using GWTC-2~\protect\cite{GWTC2TGR} (dot dashed traces) and GWTC-1~\protect\cite{GWTC1TGR} (dashed) events.  The vertical dashed line shows the GR prediction. Triangles mark the GWTC-3~\protect\cite{GWTC3TGR} medians, and vertical bars the symmetric 90\%-credible intervals. 
    } \label{fig:imrct_hier}
  \end{minipage}
\end{figure}

The second approach, the \emph{inspiral--merger--ringdown consistency test}~\cite{Ghosh:2016qgn,Ghosh:2017gfp}, checks whether the low- and high-frequency content of the underlying signal are consistent with each other. Each portion can be used to independently infer the mass and spin of the remnant object which are expected to be consistent if the entire signal is well-described by GR. Alternatively, any fractional deviation between these two estimates should be consistent with zero. This test requires a minimum SNR in each portion to perform reliable parameter inference, and we restrict the analysis to the 6 events which pass such ann SNR threshold.  The joint posterior probability distribution on the fractional deviations in final mass and spin for all the events is shown in Fig.\ref{fig:imrct_hier} along with a comparison with previous results. We do not find a violation of GR and the overall improvement in the results is upto a factor 1.8. 

\subsection{Tests of gravitational wave generation}
\label{sec:gen}

\subsubsection{Generic modifications}
\label{sec:par}

\begin{figure*}[h!tp]
\centering
\includegraphics[width=0.8\textwidth]{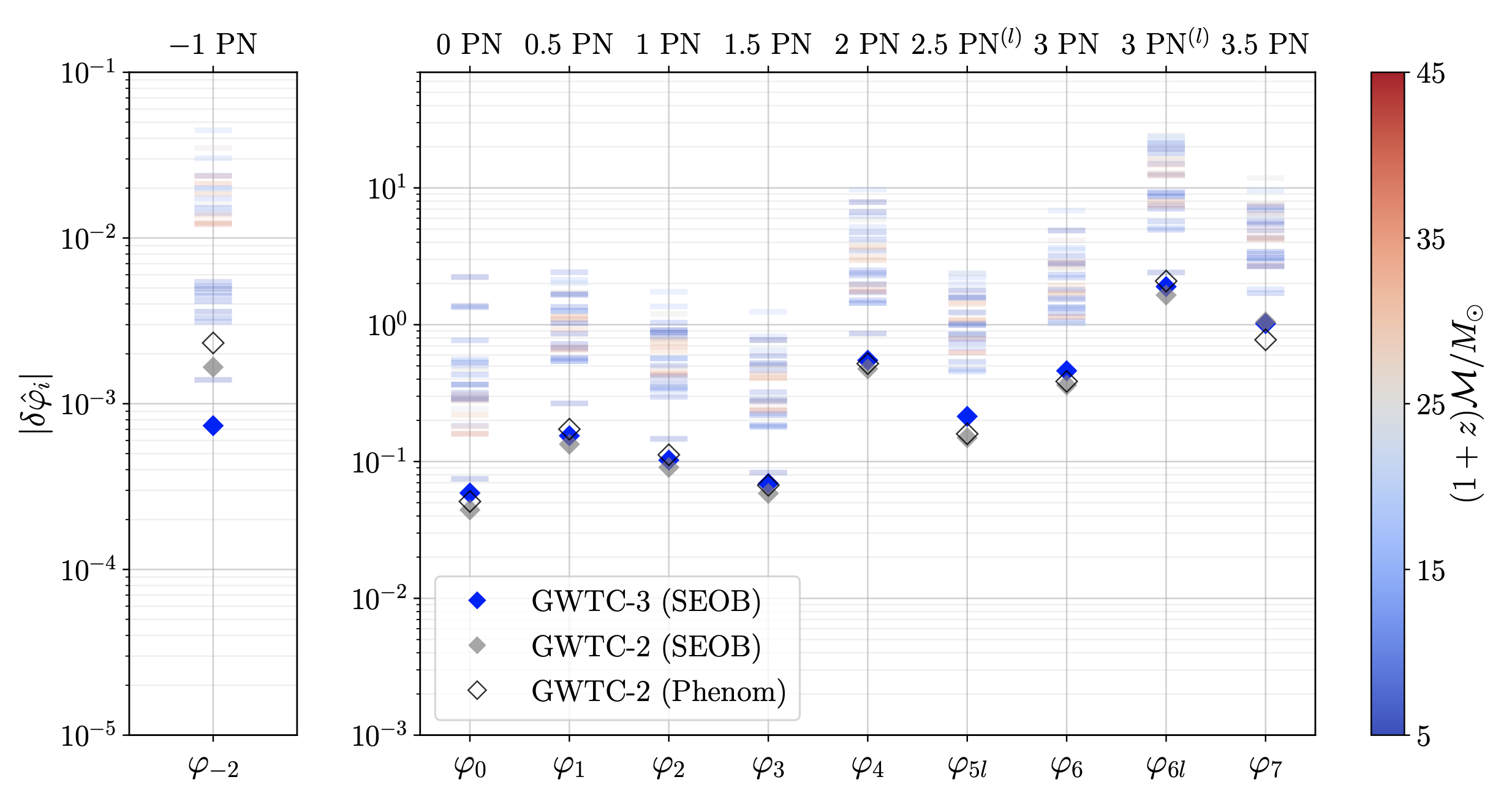}
\caption{$90\%$ upper bounds on the magnitude of the parametrised test coefficients from $-1$PN to $3.5$PN order as discussed in Sec.\,\ref{sec:par}. Bounds marked by blue diamonds were obtained with a pipeline based on a SEOB waveform combining all eligible events from O1, O2 and O3. Filled (unfilled) gray triangles mark analogous results obtained with GWTC-2 data \protect\cite{GWTC2TGR} using SEOB (\textsc{Phenom}) models.  Horizontal stripes indicate constraints obtained with individual events, with bluer (redder) colors representing lower (higher) total mass events. }
\label{fig:par:fta_bounds}
\end{figure*}    
 
Additional fields or higher-order curvature corrections introduced in alternative theories of gravity can alter the binary dynamics and leave an imprint on the GW signal.As mentioned in the introduction, a GW inspiral is best described by PN theory,  a perturbative expansion in powers of $v/c$, with each $O([v/c]^{2n})$ being referred to as of $n$PN order~\cite{Blanchet:2013haa,Buonanno:1998gg}.  Given the binary masses and spins, coefficients of each $n$PN order term is uniquely specified in GR.  The state-of-the art GR waveform models (c.f. Sec.~\ref{sec:inference}) carry terms upto the 3.5 PN order or $n=7$ and we introduce parametric deviations to these coefficients~\cite{Mishra:2010tp,Li:2011cg,Meidam:2017dgf}. Using 9 (out of 15) low-mass inspiral-dominated events, we place bounds on possible deviations of these parameters (Fig.~\ref{fig:par:fta_bounds}), which generally improve on previous results.

In GR,  an inspiralling binary does not produce dipole radiation. By introducing a phenomenological pre-Newtonian term, $\varphi_{-2}$ (Fig.~\ref{fig:par:fta_bounds}) to desribe this dipole radiation, we can bound its contribution to the binary dynamics. Although still consistent with being zero, 
\main is able to improve the bound on the measurement by a factor of $\sim 2$, the single largest improvement for any PN parameter, compared to previously published results. This improvement is dominated by the long-inspiral BH-NS candidate, GW200115\_042309.
    
\subsubsection{Spin-induced quadrupole moment}
\label{sec:sim}

A spin-induced deformation of a compact object like a BH or NS is expected to modify the GW signal.  A measurement of this physical effect can effectively help to identify the nature of the compact object and distinguish BHs or NSs from classes of ECOs~\cite{Krishnendu:2017shb}. Since, larger the spins of the individual compact objects, more pronounced this effect is  supposed to be, \main uses the twin selection criteria of low-mass \emph{and} spin measurements to select the events for this test. These events (6 out of 15) are used to make a confident measurement of possible departures of this spin-induced multipole parameter from the BH prediction, and do not find any non-BBH evidence. The  event GW191216\_213338 provides the best single-event bound. Overall we are able to restrict positive deviations more, and thus provide some meaningful bounds on certain classes of boson stars.

\subsection{Tests of gravitational wave propagation}
\label{sec:liv}

In GR, GWs propagate non-dispersively at the speed of light. There are alternate theories of gravity, like massive graviton theories, Lorentz-violating theories, etc, which predict dispersion of GWs, i.e., different frequency components of the wave travelling at different speeds~\cite{Will:1997bb,Mirshekari:2013vb}. This is expected to affect the morphology of the signal, and lead to a measurable effective de-phasing. Since this is a propagation effect, it is stronger for the more distant sources. We use a parametrised dispersion relation~\cite{Mirshekari:2011yq}:

\begin{equation}
E^2 = p^2c^2 + A_{\alpha}p^{\alpha}c^{\alpha}
\end{equation}
where the first term on the right-hand-side is the GR term and the second term aims to capture generic modifications. We place bounds on the magnitudes of dispersion $|A_{\alpha}|$ for different values of $\alpha$ and find no evidence of a deviation from GR.  The $A_0$ coefficient stands out because a bound (for $A_0> 0$) can be mapped onto a bound on the mass of the graviton, $m_g = A_0/c^2$. This bound has now been improved with respect previous analyses~\cite{GWTC2TGR}, and currently stands $2.5$ times better than the Solar System bound, although observations are still consistent with the graviton being massless.

\subsection{Polarizations}
\label{sec:pol}

In GR, GWs can have two (tensor) polarizations, the \emph{plus} (+) and the \emph{cross} ($\times$). However,  a general metric theory of gravity additionally permits two scalar and two vector modes~\cite{Eardley:1973br,Eardley:1973zuo}. A network of GW detectors allows us to probe beyond-GR polarizations. Using a linear combination of detector outputs which contains no GW signal, the \emph{null stream}, \main uses a waveform-agnostic way to check if the residuals (described earlier in Sec.~\ref{sec:con}) are consistent with our noise model for a particular assumption about the polarization content of the underlying signal. While earlier analyses~\cite{GWTC2TGR} tested hypothesis where only pure polarization states were present (either all scalar, vector or tensor), \main also investigates the possibility of mixed polarizations. Combining all eligible events across all three observing runs and computing the Bayes Factors in favour of a given hypothesis of signal polarization content~\cite{Wong:2021cmp}, we do not find evidence of beyond-GR polarization modes in the data.

\subsection{Remnant properties}
\label{sec:rem}

\subsubsection{Ringdown}
\label{sec:rin}

A consequence of the no-hair conjecture~\cite{Israel:1967wq,Carter:1971zc} is that a BBH ringdown can be uniquely described by a superposition of exponentially damped sinusoids called the quasi-normal-mode (QNM) spectrum~\cite{Vishveshwara:1970zz} whose frequencies and damping times depend only on the remnant mass and spin. Hence if one is able to measure these QNM frequencies independently, it would be a probe of the remnant object and a test of the no-hair conjecture~\cite{Berti:2005ys}. In \main two tests of BH ringdown are presented. The first one, using the \textsc{pyRing} GW data analysis toolkit~\cite{pyRing}, studies the post-merger portion of the signal and performs hypothesis testing against a continuum of models with higher modes and/or overtones. It does not find evidence for higher modes~\cite{Carullo:2019flw,Isi:2019aib} and among the events analysed (5), GW200224\_222234 shows weak evidence for overtones. Using a model where the frequency of the $\ell=2, m=2, n=1$ QNM`~\footnote{$(\ell,m,n)$ are the standard indices used to denote modes in spherical harmonics decompositions.} is allowed to deviate from its GR prediction, a marginal improvement of its bounds is found over previous analyses~\cite{GWTC2TGR}.

\begin{figure}
  \begin{minipage}[l]{0.6\textwidth}
    \includegraphics[width=\textwidth]{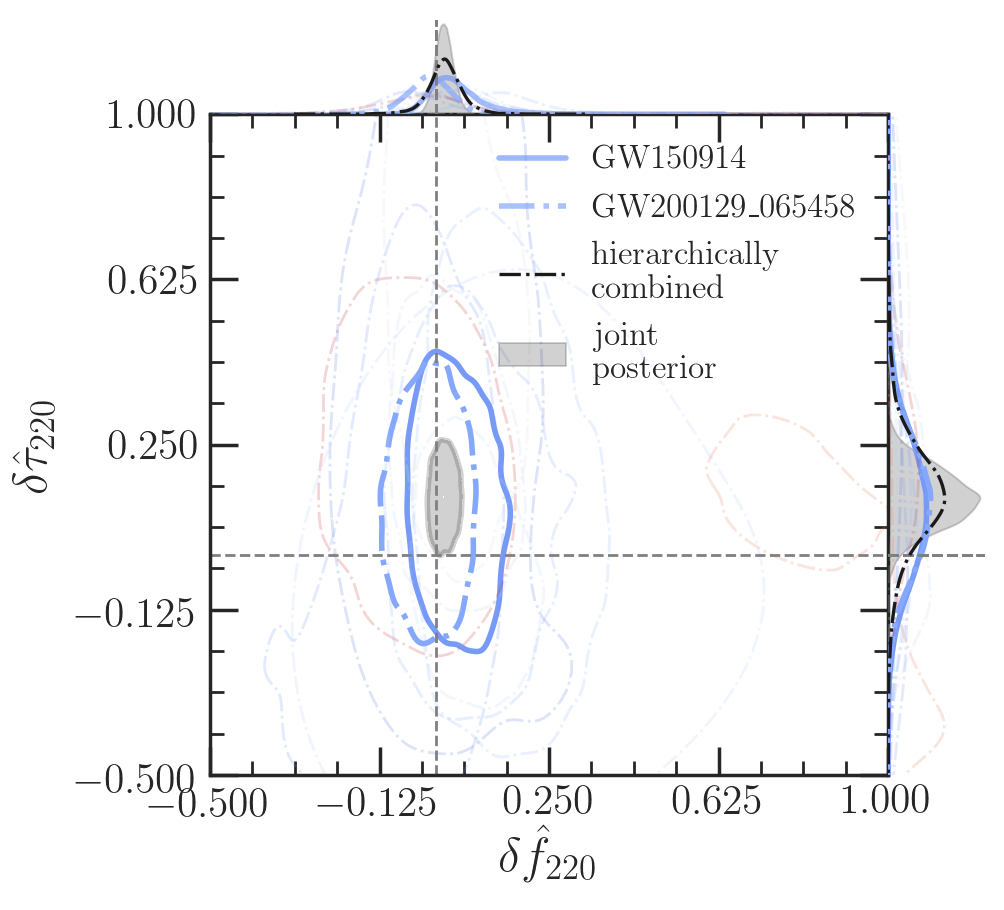}
  \end{minipage}\hfill
  \begin{minipage}[l]{0.4\textwidth}
    \caption{The 90\% credible levels of the posterior probability distribution of the fractional deviations in the frequency and damping time of the $\ell=2, m=2, n=0$ QNM, $(\delta f_{220},\delta \tau_{220})$ and their corresponding one-dimensional marginalized posterior distributions, for events from O1, O2 and O3 passing a SNR threshold of $8$ in both the pre- and post-merger signal. Posteriors for GW150914 and GW200129\_065458 are separately shown. The joint constraints on $(\delta f_{220},\delta \tau_{220})$ obtained using the restrictive method of combination from individual events are given by the filled grey contours, while the hierarchical method of combination yields the black dot dashed curves in the 1D marginalized posteriors. 
    } \label{fig:rin:qnm_deviation220pSEOBNR}
  \end{minipage}
\end{figure}

The other test of the BH ringdown makes use of a full GW inspiral-merger-ringdown (not just post-merger) waveform model to measure the $\ell=2, m=2, n=0$ QNM frequency and damping time~\cite{Ghosh:2021mrv}. In the baseline GR SEOB model with higher modes (c.f Sec.~\ref{sec:inference}), the QNM spectra is predicted from initial masses and spins using NR-inspired fitting formulae. In this test, we allow these GR predictions to deviate and constrain these deviations from the data. Like the other ringdown test, the pSEOB test focusses on high-mass events where the merger-ringdown signal is expected to be prominent. Additionally, since this test uses the entire signal,it also uses the overall SNR as a selection criterion. Fig.~\ref{fig:rin:qnm_deviation220pSEOBNR} shows results from the 6 new events from O3b using these selectionn criteria, along with combined results from previous analyses. The bounds on the fractional deviations improve by a factor 1.7-5.5 over the previous results~\cite{GWTC2TGR}. This improvement is mainly down to the the large number of events considered in \main over previous publications.
    
 \subsubsection{Echoes}
 \label{sec:ech}
 
Another way to test the BH nature of the remnant object is the search for GW echoes, repeating pulses of gravitational radiation expected to be present in the post-merger signal if the remnant is an ECO with a reflective surface instead of a classical BH with an event horizon~\cite{Cardoso:2016rao,Cardoso:2017cqb,Cardoso:2016oxy}. We search for these echoes with  \textsc{BayesWave}~\cite{Cornish:2014kda,Littenberg:2014oda} using minimal assumptions about their shapes. We compute the Bayes factors between the hypothesis that our data contains echoes verses it is just noise.  We also quantify the significance of our ``with echoes"-hypothesis using p-values for each event, similar to the residuals test from Sec.~\ref{sec:con}~\cite{Tsang:2018uie,Tsang:2019zra}. We do not find any  compelling evidence for their presence in the post-merger signal. 

\section{Conclusions}
\label{sec:conclusion}

These proceedings provide a summary of the ``Tests of General Relativity with GWTC-3" paper which includes the latest results with GW observations by the Advanced LIGO-Virgo detectors.  The paper focusses of events detected in the second half of the third observing run (O3b), but wherever possible, combines information with events from previous observing runs to give the tightest state-of-the-art bounds possible. The main paper outlines nine different methods which are used to characterise deviations from GR. There is no statistically significant evidence  found for a possible deviation  from GR.

\section{Acknowledgement}
Acknowledgements may be found in https://dcc.ligo.org/P2100218. This material is also based upon work supported by NSF’s LIGO Laboratory which is a major facility fully funded by the National Science Foundation. The author is grateful to K.G.Arun for carefully reading the manuscript and providing useful comments, and to F. Stoecker for insightful discussions during the process of writing. This document is LIGO-P2200097.

\end{document}